\documentclass[journal=jcpa,manuscript=article]{achemso}
\usepackage[version=3]{mhchem} 
\usepackage{color}
\setkeys{acs}{usetitle = true}
\def\cm{cm$^{-1}$}
\def\Sone{S$_0$$\rightarrow$S$_1\;$}
\def\Stwo{S$_0$$\rightarrow$S$_2\;$}
\newcommand*{\bra}[1]{\left<#1\right|}
\newcommand*{\ket}[1]{\left|#1\right>}
\newcommand*{\pref}[1]{Eq.~(\plainref{#1})}
\newcommand*{\fref}[1]{Fig.~\plainref{#1}}
\newcommand*{\tref}[1]{Tab.~\plainref{#1}}

\author{M. Schr\"oter}
\affiliation[Universit\"at Rostock]
{Institut f\"ur Physik, Universit\"at Rostock, D-18051 Rostock, Germany}
\author{O. K\"uhn}
\affiliation[Universit\"at Rostock]
{Institut f\"ur Physik, Universit\"at Rostock, D-18051 Rostock, Germany}
\email{oliver.kuehn@uni-rostock.de}

\title{Interplay Between Non-adiabatic Dynamics and Frenkel Exciton Transfer in Molecular Aggregates: Formulation and Application to a Perylene Bismide Model}
\begin{document}
%
\begin{abstract}
The quantum dynamics of linear molecular aggregates in the presence of \Sone and \Stwo transitions is investigated putting emphasis on the interplay between local non-adiabatic S$_2$ to S$_1$ deactivation and 
Frenkel exciton transfer. The theoretical approach combines aspects of the linear vibronic coupling and Frenkel exciton models. Dynamics calculations are performed for the absorption spectrum and the electronic state populations using the multiconfiguration time-dependent Hartree approach. As an application perylene bisimde J-type dimer and trimer aggregates are considered, including four tuning and one coupling mode per monomer. This leads to a dynamical model comprising up to 7 electronic states and 15 vibrational modes. The unknown non-adiabatic coupling strength is treated as a parameter that is chosen in accordance with available absorption spectra. This leaves some flexibility that can be limited by the clearly distinguishable population dynamics.
\end{abstract}

\section{Introduction} 
\label{intro}
Exciton-vibrational coupling (EVC) plays an important role in the dynamics and spectroscopy of natural and artificial
molecular aggregates.\cite{Kobayashi:2012wc,spano10:429,kuhn11_47,renger01:137,Pullerits:2013dt} In particular the
non-perturbative and non-Markovian regime has attracted recent interest and a number of approaches have been developed
to cope with such
situations.\cite{renger96:15654,tanimura06_062001,seibt09:13475,roden09:044909,Ritschel:2011dg,nalbach11_063040,polyutov12_21,yan12_105004}
Often the signatures of strong EVC are visible as vibrational side bands already in linear absorption or emission
spectra.\cite{wuerthner11:3376,spano11:5133,Eisfeld:2005cg} However, EVC is not the only manifestation of the
interaction between electronic and nuclear degrees of freedom. A frequent theme in the dynamics of excited electronic
states of molecules are non-adiabatic transitions signifying the breakdown of the Born-Oppenheimer
approximation.\cite{Domcke04} The topic of this contribution is the interplay between EVC in exciton transfer and local
non-adiabatic transitions.  {This issue has been intensively studied in the context of exciton-exciton
  annihilation.\cite{brueggemann01_11391} There highly excited states at energies about twice the \Sone excitation
  energy are involved. In contrast we focus on situations} where the monomers comprising the aggregate have overlapping
electronic transitions. In terms of exciton theory this will lead to the emergence of two excitonic bands being
connected by interband transitions which are due to non-adiabatic couplings.  This situation did not receive much
attention so far, although it should be rather common for many dyes. For instance, Kobayashi and coworkers have made
vibronic coupling (Herzberg-Teller coupling) between Q and B state derived exciton manifolds responsible for observed
vibrational dynamics after Q-band excitation in porphyrin type J-aggregates.  \cite{Kano:2012hz} 
 {Recently, Q- and B-band dynamics of bisporphyrin have been stduied using transient absorption and two-dimensional
spectroscopy.\cite{Kullmann:2012tf} Transient absorption data revealed time scales of associated intraband relaxation
and S$_2$-S$_1$ internal conversion in the few hundred femtosecond regime.}
 Interband transitions have also been discussed for tubular aggregates \cite{milota09:45}. However, in that case the excitonic manifolds correspond to different structural elements such as inner and outer tubes which have some residual electronic coupling. As a note of caution we stress that the present non-adiabatic coupling between different \emph{local} electronic states should not be confused with the discussion of non-adiabaticity in the context of the potential energy surfaces (PES) coupled via the Coulomb interaction between different monomers \cite{Beenken:2002jl,Eisfeld:2005cg,Tiwari:2012is}. This also holds true for studies of  molecular complexes that are not treated as an aggregate but a supermolecule. For example, Leutwyler and coworkers interpreted splittings at the electronic band origin of \Sone and \Stwo transitions of hydrogen-bonded homo-dimers in terms of excitonic coupling quenched by nuclear tunneling. \cite{ottiger12_174308}

\begin{figure}[t]
\begin{center}
\includegraphics[width=0.6\textwidth]{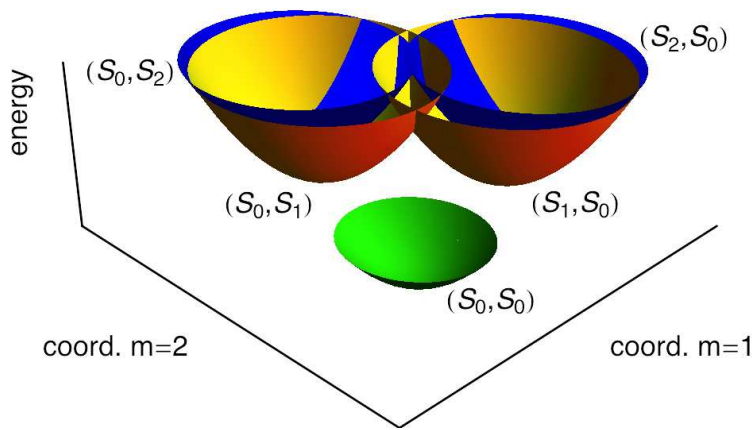}
\caption{(color online) Schematic view of the topology of the diabatic PES for a molecular dimer (site index $m=1,2$) including one vibrational coordinate and three electronic states, $S_0$-S$_2$, per monomer. The different excitation states of the dimer are indicated.}
\label{fig:scheme}     
\end{center}
\end{figure}

The situation we will focus on is illustrated in \fref{fig:scheme} for the case of a dimer with one harmonic vibrational coordinate and three electronic states for each monomer. The lowest state, (S$_0$,S$_0$), corresponds to the situation where both monomers are in their electronic ground state, S$_0$. Next in energy there are two diabatic PES corresponding to the situation where a single S$_0$-S$_1$ type excitation is present, either in monomer 1 (S$_1$,S$_0$) or 2 (S$_0$,S$_1$). For the depicted symmetric case the two PES intersect along the diagonal of the plot. In the adiabatic representation one has an avoided crossing due to the Coulomb coupling (cf. ref. \citenum{Eisfeld:2005cg}). If there is a close lying higher intramonomer electronic state, S$_2$, another set of Coulomb coupled diabatic PES will be present corresponding to the excitation states (S$_2$,S$_0$) and (S$_0$,S$_2$). Since the local shifts of the electronically excited states along the considered nuclear coordinate in general are different, the related PES will intersect. Assuming that besides the depicted tuning modes, there are some coupling modes, conical intersections (CIs) will emerge along the PES intersection lines.\cite{Domcke04}   
%
The two types of state couplings will manifest themselves in the absorption spectral lines shapes. Furthermore, one might think of interesting quantum dynamical implications. For instance, consider the limiting case where a system initially prepared in a local S$_2$ excitation will evolve either in the S$_2$ manifold or after non-adiabatic transition in the S$_1$ manifold. Which dynamics is relevant will depend on the ratio of non-adiabatic and Coulomb coupling. In fact if both are of the same magnitude intra- and interband transfer may be strongly mixed. An important aspect for the dynamics is the coupling between electronic and nuclear degrees of freedom because it influences both local non-adiabatic transitions and transfer (e.g. shifts of PES). In general, it will not be a single vibrational mode that mediates both processes. 

In the present contribution the scenario of \fref{fig:scheme} will be studied for the exemplary case of a modified  perylene bisimide (PBI) dye forming J-aggregates in solution \cite{li08:8074}. The choice of the system is motivated by our previous studies, which focussed on the quantum chemical and spectroscopic characterization of monomers and small aggregates. \cite{ambrosek11_17649,Ambrosek:2012ek} Here we will extend our model to include the S$_2$ state as well as non-adiabatic S$_2$$\rightarrow$S$_1$ transitions. In the following section we start with outlining the theoretical approach which combines the Frenkel exciton Hamiltonian with vibronic coupling theory in the diabatic representation. Next some details of the quantum dynamics method for calculating absorption spectra and population dynamics are given. This section concludes with a summary of the model parameters. The Results section focusses on the spectral features due to the different couplings as well as on the corresponding regimes of population dynamics.
\section{Theoretical Methods}
\label{sec:methods}
\subsection{Frenkel-Exciton Vibronic-Coupling Hamiltonian} 
In the following we will give a short summary of the Frenkel exciton Hamiltonian with special emphasis on the vibronic coupling between different local diabatic electronic states, $\ket{\varphi_{m,a}}$ ($m=1\ldots N_{\rm agg}$ is the site index and $a=0,1,2$ counts the electronic states).  The aggregate Hamiltonian can be written as \cite{may11}
\begin{eqnarray}
H_{\rm agg}=\sum_{m}\sum_{ab} H_m(a,b) |\varphi_{m,a}\rangle \langle \varphi_{m,b}| 
+ \frac{1}{2}\sum_{mn}\sum_{abcd}J_{mn}(a_mb_n,c_nd_m) |\varphi_{m,a}\varphi_{n,b}\rangle \langle \varphi_{n,c}\varphi_{m,d}| \, .
\label{eq:hagg}	
\end{eqnarray}
Within the shifted harmonic oscillator model the diagonal on-site part is given by ($E_{m,a}$: electronic energy)
\begin{eqnarray}
H_{m}(a,a) &=& U_{m,a}+\sum_\xi\frac{\hbar\omega_{m,a}(\xi)}{2} \left( - \frac{\partial^2}{\partial Q_{m,\xi}^2}+ Q_{m,\xi}^2\right) \, ,\\
U_{m,a} & = & E_{m,a}    +\sum_\xi\hbar \omega_{m,a}(\xi) g_{m,a}(\xi)Q_{m,\xi} \, .
\end{eqnarray}
Here $\{ Q_{m,\xi} \}$ denotes the set of dimensionless electronic ground state normal modes at site $m$, $\omega_{m,a}(\xi)$ are the respective frequencies, and $g_{m,a}(\xi)$ is the linear coupling in state $\ket{\varphi_{m,a}}$ ($g_{m,a}(\xi) = (\partial U_{m,a}/\partial Q_{m,\xi})/\hbar \omega_{m,a}(\xi) $).
The coupling of a particular mode to the electronic transition can be characterized by the (dimensionless) Huang-Rhys factor $S_{m,a}(\xi) =  g_{m,a}^2(\xi)/2$.
These modes are also called tuning modes since they influence the energy gap between different electronic states. 

For the off-diagonal elements a linear expansion in terms of the set of coupling modes $\{ Q_{m,\eta} \}$ is assumed to be valid (linear vibronic coupling (LVC) model), i.e. \cite{koppel84:59}
\begin{eqnarray}
H_{m}(a,b) =  \sum_\eta  \lambda_{m,\eta}(a,b) Q_{m,\eta}
\end{eqnarray}
with $\lambda_{m,\eta}(a,b)$ being the respective coupling matrix. The $N_{\rm vib}$ modes of the $m$th site will be comprised into the vector $\mathbf{Q}_m=(\{Q_{m,\xi}\},\{Q_{m,\eta}\})$

Electronic transitions at different sites are coupled via the Coulomb matrix element $J_{mn}(a_mb_n,c_nd_m)$ in  \pref{eq:hagg}. 
In the following we will assume that concerning the Coulomb coupling only close to resonant interactions involving ground state transitions need to be accounted for, i.e. the only non-zero elements of $J_{mn}(a_mb_n,c_nd_m)$ are $J_{mn}(1_m0_n,1_n0_m)=J_{mn}(0_m1_n,0_n1_m)$ and  $J_{mn}(2_m0_n,2_n0_m)=J_{mn}(0_m2_n,0_n2_m)$.
Further,  $J_{mn}$ will be assumed to be independent of the nuclear coordinates.

Electronic excitations are usually classified according to zero-, one-, two- etc. excitation states. Restricting to the one-exciton space we have the following completeness relation 
%
\begin{equation}
	\mathbf{1} = \ket{0}\bra{0} + \sum_{m}\sum_{a=1,2} \ket{m_a}\bra{m_a} \, ,
\end{equation}
with 
\begin{equation}
	\label{eq:basis}
	  \ket{0}=\prod_m \ket{\varphi_{m,0}}, \quad \quad \ket{m_a} = \ket{\varphi_{m,a}} \prod_{k\ne m} \ket{\varphi_{k,0}} \, . 
\end{equation}
Thus we have the correspondences: $\ket{0} \rightarrow$ (S$_0$, S$_0$, $\ldots$), $\ket{1_a} \rightarrow$ (S$_a$, S$_0$, $\ldots$) and so on.
The aggregate Hamiltonian expressed in this basis can be written as 
\begin{eqnarray}
	H_{\rm agg} &=& H^{(0)} + H^{(1)} \, , \\
	H^{(0)} &=&  \sum_m H_m(0,0) \ket{0}\bra{0} \equiv \mathcal{E}_0\ket{0}\bra{0} \, ,\\
	H^{(1)} & = &  \sum_{mn}\sum_{a,b=1,2}[ \delta_{mn}(\delta_{ab}(\mathcal{E}_0 + U_{m,a}) + (1-\delta_{ab})H_m(a,b)) \nonumber\\
	&+&  J_{mn}(a_m,0_n,b_n,0_m)] |m_a\rangle \langle n_b| \, .
\end{eqnarray}
%
%
\subsection{Quantum Dynamics}
The time-dependent Schr\"odinger equation will be solved employing the multiconfiguration time-dependent Hartree (MCTDH) method.\cite{meyer90:73,beck00:1} To this end the state vector is expanded in terms of the diabatic basis, i.e
\begin{eqnarray}
	|\Psi({\bf Q};t) \rangle=\sum_{\alpha} \chi_{\alpha}({\bf Q};t) \, |\alpha\rangle	\, ,
\end{eqnarray}
where $\alpha$ runs over all electronic configurations (cf. \pref{eq:basis}) and the nuclear coordinates are comprised in the $D=N_{\rm agg}\times N_{\rm vib}$ dimensional vector $\mathbf{ Q}=(\mathbf{Q}_1, \ldots, \mathbf{Q}_{N_{\rm agg}})\equiv (Q_1,\ldots,Q_D)$.

The nuclear wave function for each diabatic state is  expanded into MCTDH form as follows
\begin{equation}
\label{eq:psiMCTDH}
\chi_\alpha(\mathbf{ Q},t) = \sum_{j_1 \ldots j_D}^{{n_{j_1} \ldots n_{j_D}}}
C^{(\alpha)}_{j_1,\ldots,j_D}(t) \phi^{(\alpha)}_{j_1}(Q_1;t) \ldots \phi^{(\alpha)}_{j_D}(Q_{D};t) \, .
\end{equation}
Here, the $C^{(\alpha)}_{j_1,\ldots,j_D}(t)$ are the time-dependent expansion coefficients weighting the contributions of the different Hartree products, which are composed of single particle functions (SPFs), $\phi^{(\alpha)}_{j_k}(Q_k;t)$, for the $k$th degree of freedom in state $\alpha$.

The aggregate models will be  characterized by means of the absorption spectrum. Here, a time-dependent formulation will
be used, i.e. the absorption spectrum at $T=0$ K is expressed as \cite{may11} ($I_{0}$ normalization constant)
\begin{equation}
\label{eq:abs}
I(\omega)=I_0 \omega \, \sum_{i=x,y}{\rm Re} \int_0^\infty dt\, e^{i\omega t-(t/\tau_i)^2} \langle \Psi_0 |d^{(i)} e^{-iH_{\rm agg}t/\hbar} d^{(i)} | \Psi_0 \rangle \, ,
\end{equation}
where $d^{(i)}$ is the $i$-th directional component of the dipole operator. For the latter we assume the Condon approximation with constant \Sone and \Stwo transition dipole moments $d_{m,a=1,2}$ to be valid. Notice that according to ref. \citenum{ambrosek11_17649} the \Sone and the \Stwo transitions have different polarization, i.e.
\begin{eqnarray}
	\label{eq:dip}
	d^{(x)} &=& \sum_m (d_{m,1} |m_1\rangle \langle 0| + {\rm h.c.}) \, , \\
	d^{(y)} &=& \sum_m (d_{m,2} |m_2\rangle \langle 0| + {\rm h.c.})\, .
\end{eqnarray}
Further, in \pref{eq:abs} $\tau_i$ is a parameter mimicking the finite line width (dephasing time) of the real
system.  {Note that we will use different values for the \Sone and \Stwo transitions in order to obtain
  better agreement with experiment. This is motivated by the fact that in contrast to the orbitals involved in the \Sone
transition those of the \Stwo transitions are partly localized on the phenyl bay substituents
(cf. ref. \citenum{ambrosek11_17649}). Their floppiness will cause stronger fluctuations of the transition energy in a
solvent environment.}

\subsection{Computational Methods} 
The setup for the electronic structure calculations has been detailed in refs. \citenum{ambrosek11_17649} and \citenum{Ambrosek:2012ek}. In brief the gas phase ground state equilibrium geometry of the PBI monomer has been obtained using density functional theory (DFT) with the B3LYP functional and a  6-311G* split-valence basis set. The first two electronically excited singlet states were calculated employing time-dependent DFT (TDDFT). 
Harmonic analysis of the ground state vibrations and projection of the forces at the vertical transitions have been used to obtain frequencies and Huang-Rhys factors. 
All quantum chemistry  calculations have been performed with the  TURBOMOLE program package. \cite{TURBOMOLE} 
For the Coulomb coupling between the \Sone transitions we have used a value which is in accord with the previous
calculation \cite{Ambrosek:2012ek} and an estimate from experiment  {(see below)}. \cite{li08:8074} For the \Stwo transitions
we have scaled the \Sone value according to the different calculated transition dipole moments assuming the dipole approximation for simplicity. 

Wave packet propagations have been performed using the Heidelberg program package \cite{mctdh84}. Three systems have
been considered, i.e. monomer, dimer, and trimer having 3, 5, and 7 diabatic states, respectively. For each monomer five
nuclear degrees of freedom are included into the model with parameters to be specified below. Four of these,
$Q_{m,1}$-$Q_{m,4}$, act as tuning modes for the system. The fifth mode, $Q_{m,c}$, represents a coupling mode. As
primitive basis we have chosen a 35 point harmonic oscillator discrete variable representation in the interval
$[-7.5:7.5]$ and $[-6.5:6.5]$ for the low- and high-frequency tuning mode, respectively.  For the  coupling mode we have
used 35 points in the interval $[-6:6]$.  For the single SPF basis the multi set method was used
(cf. \pref{eq:psiMCTDH}). Further, we employed mode combination with respect to the tuning modes, i.e. alike modes of
the different monomers $Q_{m,\xi}$ ($m=1,\ldots,N_{\rm agg}$; $\xi=1,...4$) have been combined to four $N_{\rm
  agg}$-dimensional sets of basis functions. As convergence criteria for the SPF basis we used that the maximal
population of the least occupied SPF should be below $10^{-3}$ in the monomer and dimer case and below $5\cdot10^{-3}$
in the trimer case.  {This required from 9/2 to 25/6 SPFs  per state and combined tuning/coupling modes for the weak and
strong coupling case, respectively.}

 {Absorption spectra have been calculated via two independent wave packet propagations according to the sum in
\pref{eq:abs} using the transition specific dipole operator in \pref{eq:dip}.} Diabatic state population dynamics is investigated starting from two different initial conditions: (i) excitation from the S$_0$ to \emph{all} S$_2$ states according to the dipole operator $d^{(y)}$ and (ii) excitation of the first monomer $m=1$ only. The latter condition, although being artificial in terms of experimental realization, is better suited for investigation of the interplay between local non-adiabatic transitions and energy transfer.
\section{Results} 
\label{sec:results}

\subsection{Application to PBI} 
\label{sub:application_to_pbi}
The electronic parameters have been adopted such as to present the situation in the J-aggregate forming PBI dye (see, refs. \citenum{li08:8074,ambrosek11_17649,Ambrosek:2012ek}). Thereby we have neglected a possible heterogeneity and assumed that all monomers have identical properties and  Coulomb couplings between neighboring monomers are the same throughout the aggregate. Specifically, we have chosen $E_{m,1}=E_1=2.13$ eV and $E_{m,2}=E_2=2.74$ eV. For the Coulomb couplings we use the following values: $J_{mn}(1_m0_n,1_n0_m)=-500$ \cm{} and $J_{mn}(2_m0_n,2_n0_m)=-150$ \cm. Since we will show normalized spectra, only the ratio between the transition dipole moments is of relevance. According to ref. \citenum{ambrosek11_17649} this ratio is given by $d_{m,1}/d_{m,2}=1.85$.

\begin{figure}[t]
\begin{center}
\includegraphics[width=0.5\textwidth]{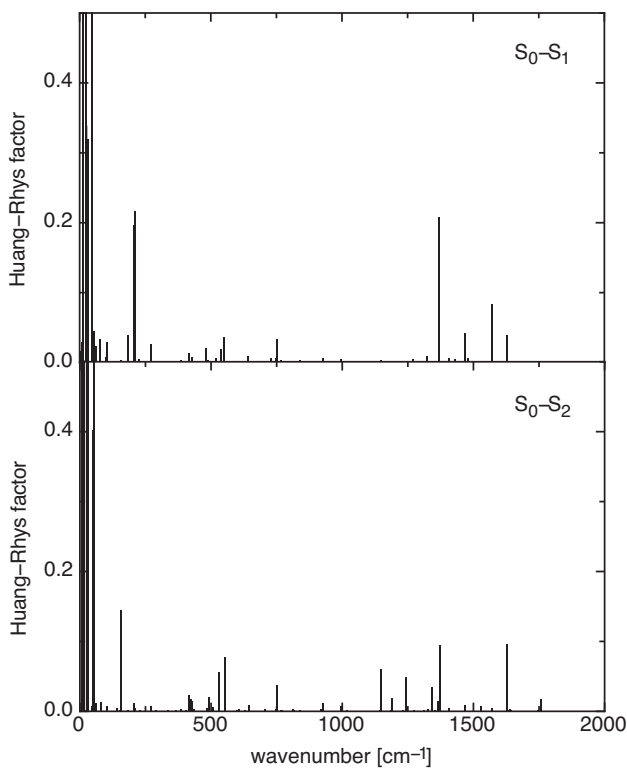}
\caption{Huang-Rhys factors for the \Sone and \Stwo electronic transitions of a PBI monomer. Notice that the low-frequency part has been cut-off since for modes in this region the harmonic approximation is less reliable.}
\label{fig:HR}     
\end{center}
\end{figure}
\begin{figure}[t]
\begin{center}
\includegraphics[width=0.37\textwidth]{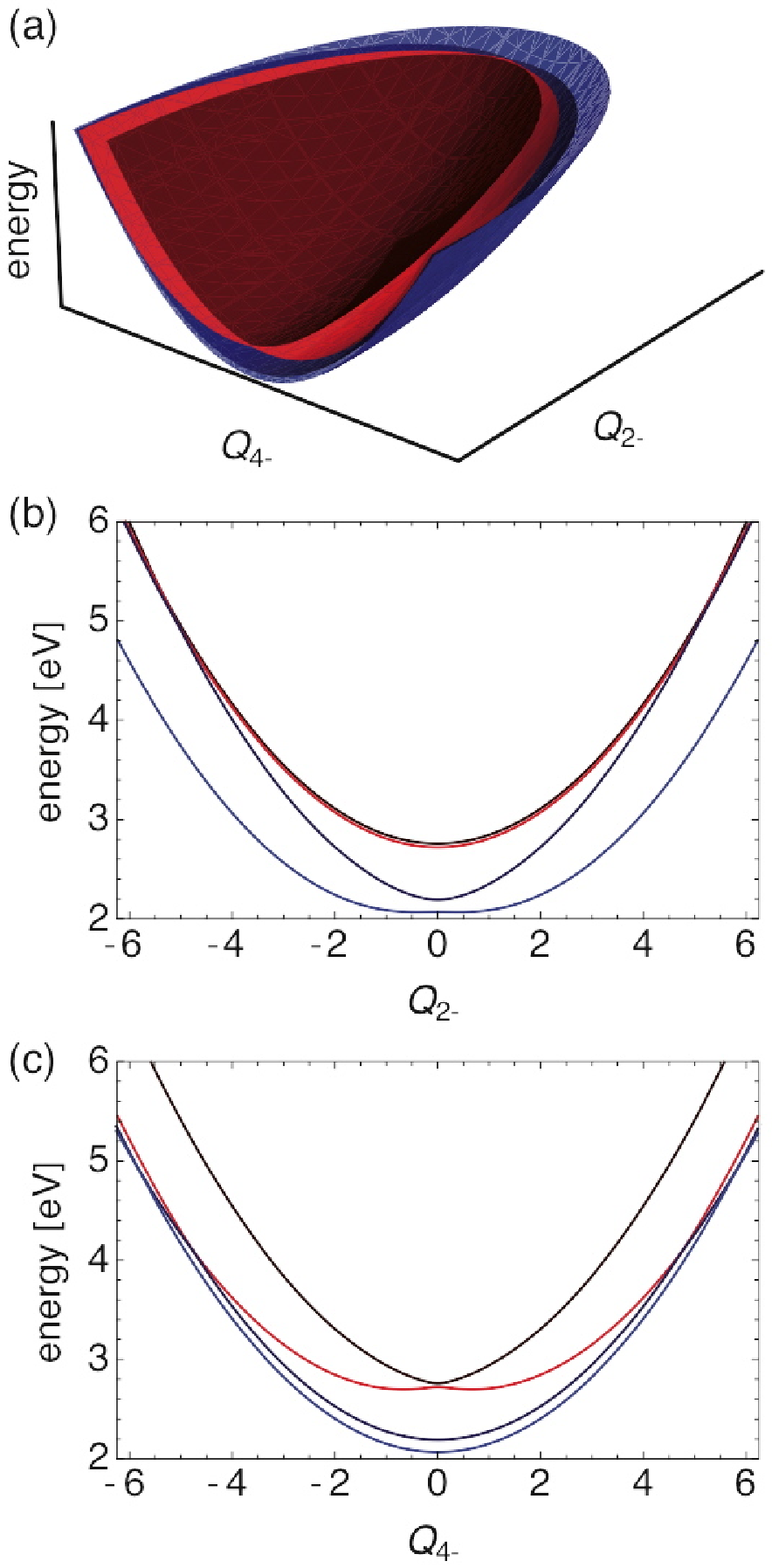}
\caption{(color online) (a) Adiabatic PES of the S$_1$ (blue) and S$_2$ (red) derived states for the dimer along antisymmetric combined modes $Q_{\xi-}=(Q_{1,\xi}-Q_{2,\xi})/\sqrt{2}$. CI's exist between the lower state of the S$_2$ band and the higher state of the S$_1$ band for non-vanishing coupling matrix element within the LVC model. Panels (b) and (c) show a cut along the $Q_{2-}$ and $Q_{4-}$ mode, respectively, for better visibility of the energetics. } 
\label{fig:PES}
\end{center}
\end{figure}

In \fref{fig:HR} we present the Huang-Rhys factors for the monomeric \Sone and \Stwo tran\-sitions. First we notice that the contributions in the low-frequency region (below 100 \cm) are of rather high magnitude. Given the potential problems of this spectral region, e.g. with respect to necessary anharmonic corrections, we will not consider frequencies below 100 \cm. In order to set up a simple model, which is extendable to larger aggregates, it will be necessary to comprise the many modes with appreciable Huang-Rhys factors into some effective modes. \cite{gisslen09:115309,ambrosek11_17649} In ref. \citenum{ambrosek11_17649} we could obtain a reasonable fit to the linear \Sone absorption spectrum by combining all modes above 800 \cm{} into one effective mode. Here we will introduce a second effective mode for the range between 100 and 800 \cm. The resulting effective mode parameters for both transitions are compiled in Tab. \plainref{tab:EFF}. From \tref{tab:EFF} we notice that in accordance with \fref{fig:HR} the two excited state configurations have nuclear displacements in different directions.
\begin{table}[t]
  \centering \begin{tabular}{|c||c|c|c|}
\hline
  mode ($\xi$) & $\omega_{m,a=1,2}(\xi)/2\pi c$  &  $S_{m,1}(\xi)$&   $S_{m,2}(\xi)$ \\
  \hline\hline
1   & 297 & 0.71  &  0\\
2   & 1416 & 0.44 &  0 \\
3   & 400  & 0 & 0.49 \\
4    &1354 & 0 & 0.47 \\
\hline
\end{tabular}
  \caption{Effective mode parameters which are calculated from the full set of modes (denoted by a tilde) as follows: frequency $\omega_{m,a}(\xi)=1/S_{m,a}(\xi){\sum_\zeta}'\tilde{S}_{m,a}(\zeta)\tilde{\omega}_{m}(\zeta)$ (given in \cm), Huang-Rhys factor $S_{m,a}(\xi)={\sum_\zeta}'\tilde{S}_{m,a}(\zeta)$, where the prime should remind on the definition of the frequency intervals for the two effective modes (see text). Note that no excited state frequencies are calculated in the original full set, i.e. only information about the vertical transition geometry is included.
}\label{tab:EFF}
\end{table}

In the following we will consider three different models, i.e. the cases $N_{\rm agg}=1,2,3$. In each case the monomer
is described by three electronic states, four tuning, and one coupling mode. Concerning the tuning modes we have used
the effective mode model of Tab. \plainref{tab:EFF}. Since there is no information with respect to the coupling mode we
have assumed that its frequency is 1000 \cm{} and the LVC constant $\lambda$ is treated as an adjustable parameter.  {Note that the qualitative behaviour is not much influenced by the actual choice of the frequency as
has been confirmed for a value of 100 \cm.}  In
order to have some guidance for the choice of $\lambda$  we will compare linear absorption spectra with experimental
results.\cite{ambrosek11_17649,marciniak11:648} Since this allows for some ambiguity we will further inspect the
population dynamics of diabatic states, which in principle could be obtained, e.g., by fitting from pump-probe spectra
\cite{kuhn11_47}.

In order to gain some insight into the topology of the PES we have plotted in \fref{fig:PES} different cuts for a dimer in adiabatic representation and using the antisymmetric combination of monomer coordinates, $Q_{\xi-}=(Q_{1,\xi}-Q_{2,\xi})/\sqrt{2}$. Panel (a) shows a global view whereas panel (b) and (c) present a cut along that high-frequency mode which is displaced  in the S$_1$ and S$_2$ state, respectively. Important  for the subsequent discussion are the Coulomb coupling induced splittings around $Q_{\xi-}=0$ and the region of approach of S$_1$ and S$_2$ surfaces for $|Q_{\xi-}|> 4$. Here, CIs emerge due to the LVC when moving along the orthogonal mode $Q_{m,c}$. The energetic position of the CI depends mainly on the difference of Huang-Rhys-factors for the S$_1$ and S$_2$ states, but is also influenced by the band splitting due to the Coulomb coupling. It should be noted, that the initial dynamics after photo-excitation proceeds not along this coordinate, but along the symmetric combination $Q_{\xi+}=(Q_{1,\xi}+Q_{2,\xi})/\sqrt{2}$. Hence, for the dimer reaching the CI requires a backscattering from the repulsive potential hit upon moving along $Q_{\xi+}$ (cf. \fref{fig:scheme}).

\begin{figure}[t]
\begin{center}
\includegraphics[width=0.70\textwidth]{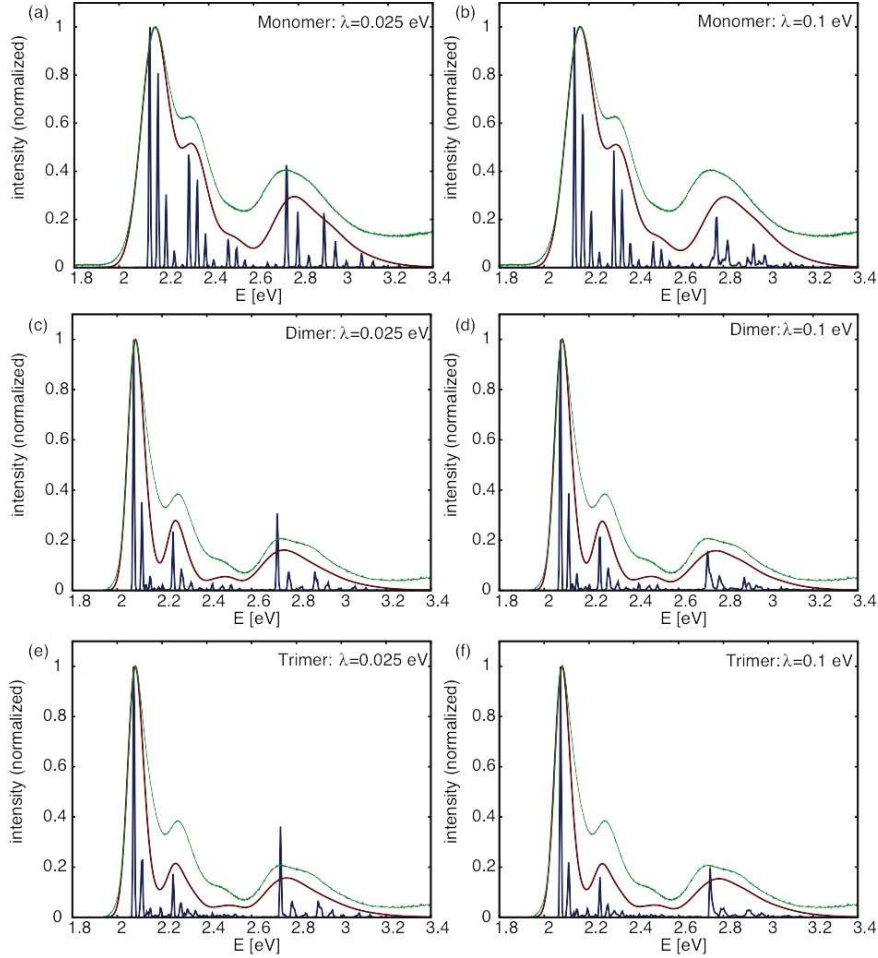}
\caption{(color online) Absorption spectra of monomer (a,b), dimer (c,d), and trimer (e,f) for two different values of the LVC constant $\lambda$. The comparison with the experimental data (ref. \citenum{ambrosek11_17649,marciniak11:648}, green line) is made using two different resolutions (blue line $\tau_x=30$ fs (\Sone) and 11 fs (\Stwo), red line $\tau_y=400$ fs). }
\label{fig:abs}     
\end{center}
\end{figure}
\subsection{Absorption Spectra} 
\label{sub:absorption_spectra}
In Fig. \plainref{fig:abs} we compare the absorption spectra of the different systems for two values of the LVC
strength $\lambda$ with experimental data obtained for monomeric \cite{ambrosek11_17649} and aggregated
\cite{marciniak11:648} PBI.  {The agreement is fairly good, given the approximate nature of the present
 reduced dimensionality
model as well as possible errors due to the quantum chemical method. } Further note that in experiment the aggregates
are most likely longer than the trimer considered here, although it has been argued that the effective coherence length
of the exciton is two monomers only; see also discussion in ref. \citenum{Ambrosek:2012ek}. The effect of J-aggregation
on the spectrum is seen, independent of the coupling strength $\lambda$, i.e. the red-most band of the spectrum increases in intensity relative to the higher energetic part. 
Because of the higher Coulomb coupling strength  between the S$_1$ states as compared with the S$_2$ states the red shift is stronger in the \Sone band than in the \Stwo band of the spectra.
\begin{figure}[t]
\begin{center}
\includegraphics[width=0.69\textwidth]{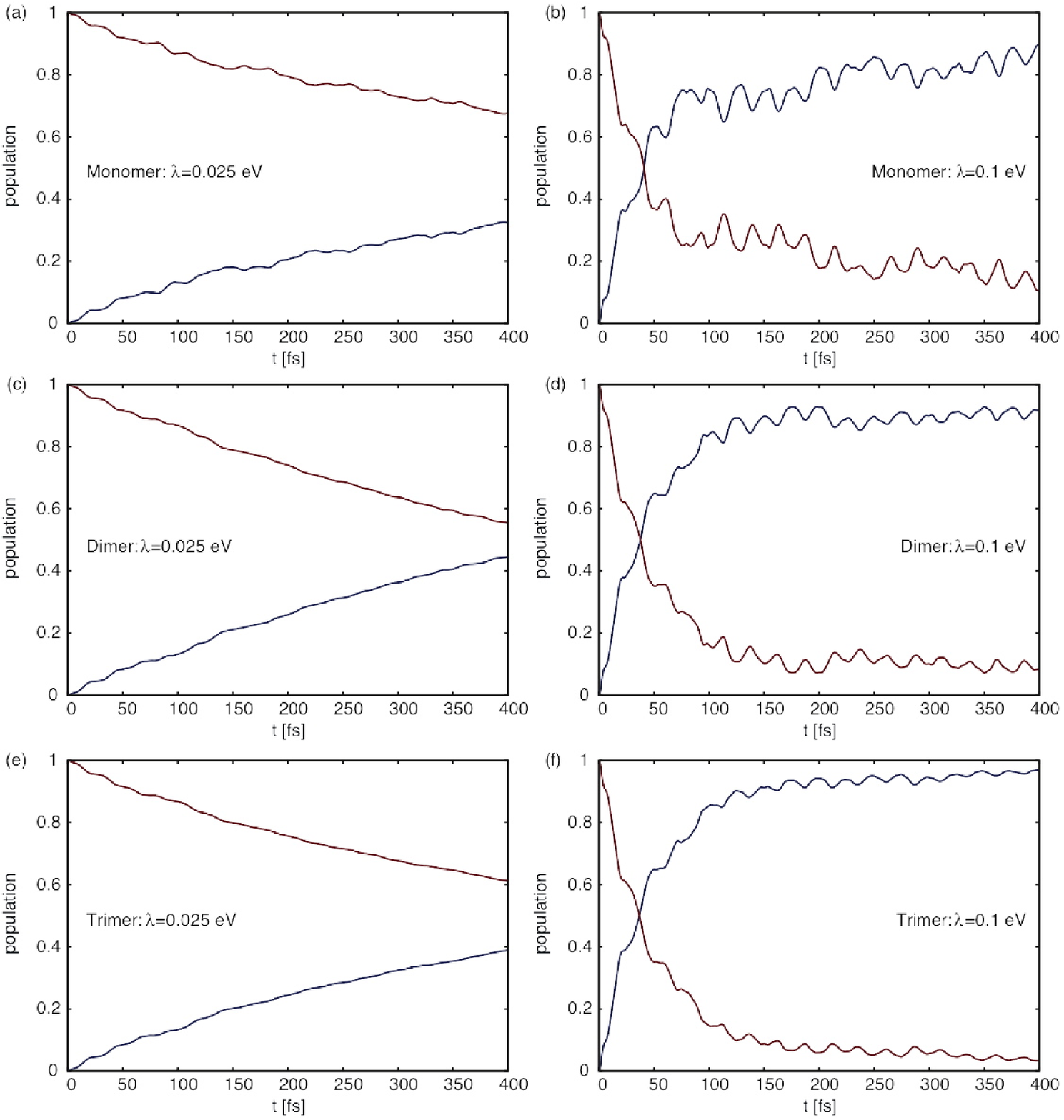}
\caption{(color online) Diabatic population dynamics of S$_2$ (red) and S$_1$ (blue) states for  monomer (a,b), dimer
  (c,d), and trimer (e,f) and for two different values of the LVC constant $\lambda$. The initial excitation is to the
  collective S$_2$ band according to the dipole operator, \pref{eq:dip}.  {For the aggregate case the populations of the
  different locally excited diabatic states have been summed. }Notice that due to symmetry all alike local excitation states behave in the same way. 
}
\label{fig:popdyn}     
\end{center}
\end{figure}
  As expected this effect of concentration of oscillator strength is more pronounced for the trimer as compared with the
  dimer. Although the experimental spectrum is considerably broadened it is instructive to inspect a high-resolution
  simulation, which gives insight into the underlying transitions. For the case of the monomer (panels (a) and (b)) the
  \Sone and \Stwo derived Franck-Condon progression are clearly discernible. They partly overlap in the range of the
  \Stwo 0-0 transition around 2.7 eV. The effect of increasing the LVC constant is most notable in the \Stwo band
  although some oscillator strength redistribution is observed even for the low-energy part (2.1-2.3 eV) of the \Sone
  band. In case of the dimer (\fref{fig:abs}(c,d)) the \Sone part of the spectrum is less affected by the LVC and in
  fact the effect of increasing the aggregate to a trimer (\fref{fig:abs}(e,f)) has a more pronounced influence. Clearly the \Sone derived band is shaped by the Coulomb coupling. In contrast for the \Stwo band both LVC and Coulomb couplings play a role. Besides the mere magnitude of the couplings it is the energetic position of the CI which is responsible for this behavior (cf. \fref{fig:PES}). Most important in this respect are those high-frequency modes with a large Huang-Rhys-factor for the S$_2$ states, which  give rise to low lying CI's.

Summarizing \fref{fig:abs} it can be stated that (i) given the approximations involved the experimental monomer and
aggregate spectra are reasonably reproduced and (ii) based on these spectra alone, the magnitude of the LVC strength
cannot be unambiguously determined.  {We note, however, that value beyond the used 0.1 eV lead to markedly
different absorption spectra and therefore can be excluded.}

\subsection{Population Dynamics} 
\label{sub:population_dynamics}
Whereas the low-resolution spectra do not show a pronounced dependence on the LVC strength over a reasonable parameter range, the population dynamics is, of course, rather sensitive to this parameter. This can be seen by comparing the left and right columns of  \fref{fig:popdyn}. For the larger value $\lambda=0.1$ eV the S$_2$ to S$_1$ interband population transfer is almost complete within 400 fs for all cases. We further notice that oscillations are more pronounced for $\lambda=0.1$ eV whereas for $\lambda=0.025$ eV the population decay appears to be almost exponential. 
Comparing the monomer with the two aggregates we observe a non-monotonic dependence on aggregate size for the smaller LVC. Here the interband decay becomes faster when going from the  monomer to the dimer, but slows down  if the aggregate is increased to a trimer. In contrast to the weaker LVC the final population after 400 fs decreases with aggregate size for the stronger LVC.

\begin{figure}[t]
\begin{center}
\includegraphics[width=0.30\textwidth]{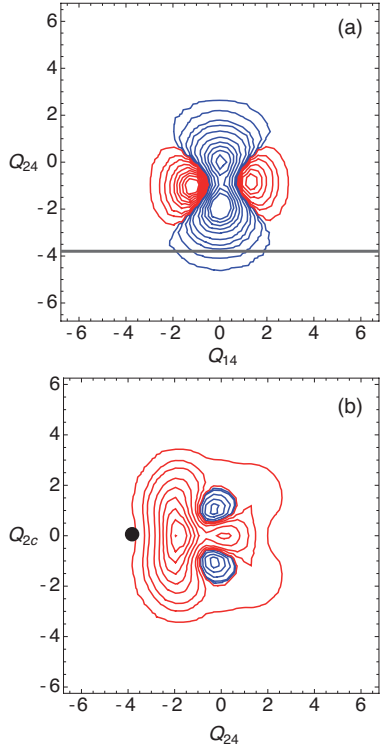}
\caption{Wave packet dynamics in the  S$_2$ state of monomer 2 of the dimer model according to \fref{fig:popdyn}
  averaged over a time interval of 400 fs.  (a)  Difference of diabatic densities \pref{eq:1} between simulation with ($|\Psi_{\rm wJ}|^2$) and without ($|\Psi_{\rm woJ}|^2$) Coulomb coupling ($\lambda=0.025$eV), (red: $|\Psi_{\rm wJ}|^2>|\Psi_{\rm woJ}|^2$, blue: $|\Psi_{\rm wJ}|^2<|\Psi_{\rm woJ}|^2$). The horizontal line shows the crossing seam between S$_2$ and S$_1$ states.
(b) Difference of diabatic densities  between simulation with small ($|\Psi_{\lambda={\rm 0.025eV}}|^2$) and large ($|\Psi_{\lambda={\rm 0.1eV}}|^2$) LVC  (red: $|\Psi_{\lambda={\rm 0.1eV}}|^2>|\Psi_{\lambda={\rm 0.025eV}}|^2$, blue: $|\Psi_{\lambda={\rm 0.1eV}}|^2<|\Psi_{\lambda={\rm 0.025eV}}|^2$). The black dot marks the CI geometry. 
(contours in interval $[-1.2:1.2]\times 10^{-5}$ (a) and $[-7:1.5]\times 10^{-6}$ (b) are chosen such as to promote visual clarity; wave functions are normalized to compensate for the population decay)
}
\label{fig:wpdiff}        
\end{center}
\end{figure}
The observed behavior can be rationalized as being the net outcome of two opposite effects. First, the energetic
position of the CI decreases with increasing aggregate size due to the influence of the Coulomb coupling (see
\fref{fig:PES}). Since this leads to a better resonance at the vertical transition energy of the wave packet, the
population transfer becomes faster as seen in the dimer case. However, as noted before the wave packet reaches the CI
only after being backscattered from the repulsive wall in $Q_{refxi+}$ direction. This is in contrast to the monomer case
where already the initial motion of the wave packet would be towards the CI. The broadening of the wave packet
associated with the backscattering causes a less effective interband transfer in the weak LVC case.  {The effect of
backscattering is quantified in \fref{fig:wpdiff}a where we plot the time-averaged reduced density differences 
\begin{equation}
  \label{eq:1}
  \Delta (Q_i,Q_j) = \int_0^T dt \int d \mathbf{Q}' \; (|\Psi_{\rm wJ}(\mathbf{Q},t)|^2-|\Psi_{\rm
    woJ}(\mathbf{Q},t)|^2)\, ,
\end{equation}
where $Q_i$ and $Q_j$ denote selected coordinates, $d \mathbf{Q}' $ includes all other coordinates and the index ``wJ''
and ``woJ'' indicates a propagation with and without Coulomb interaction, respectively, up to final time $T$.}

Clearly, the presence of the Coulomb coupling leads on average to less density in the range of the crossing seam.  In the dimer case the shift of the CI still dominates the effect of wave packet spreading, making the interband transfer faster as compared with the monomer. For the trimer however, the increased number of vibrational degrees of freedom allows for a pronounced intraband redistribution of the wave packet on the S$_2$ manifold what slows down the interband decay.

Finally, we comment on the observation of oscillations in the strong LVC case in \fref{fig:popdyn}. In case of weak
coupling there is some continuous but relatively small leakage of density out of the S$_2$ states, whenever the wave
packet hits the crossing region. In contrast for strong LVC the S$_2$ population drops by about 40 \% already on the
first attempt. The subsequent oscillations are rather a signature of quantum interferences of parts the wave packet
running on different states and being projected onto the diabatic electronic basis. The situation is visualized in
\fref{fig:wpdiff}b where we show the time-averaged density differences along two selected modes for the cases of small
and large LVC in the dimer model. The wider spread of density for the larger LVC is clearly discernible.  {Finally, it
should be noticed that these oscillations are likely to disappear upon increasing the dimensionality of the model to
approach a quasi-continuous spectrum of exciton-vibrational states. }
\subsection{Deactivation versus Transfer} 
\label{sub:deactivation_versus_transfer}
\begin{figure}[t]
\begin{center}
\includegraphics[width=0.82\textwidth]{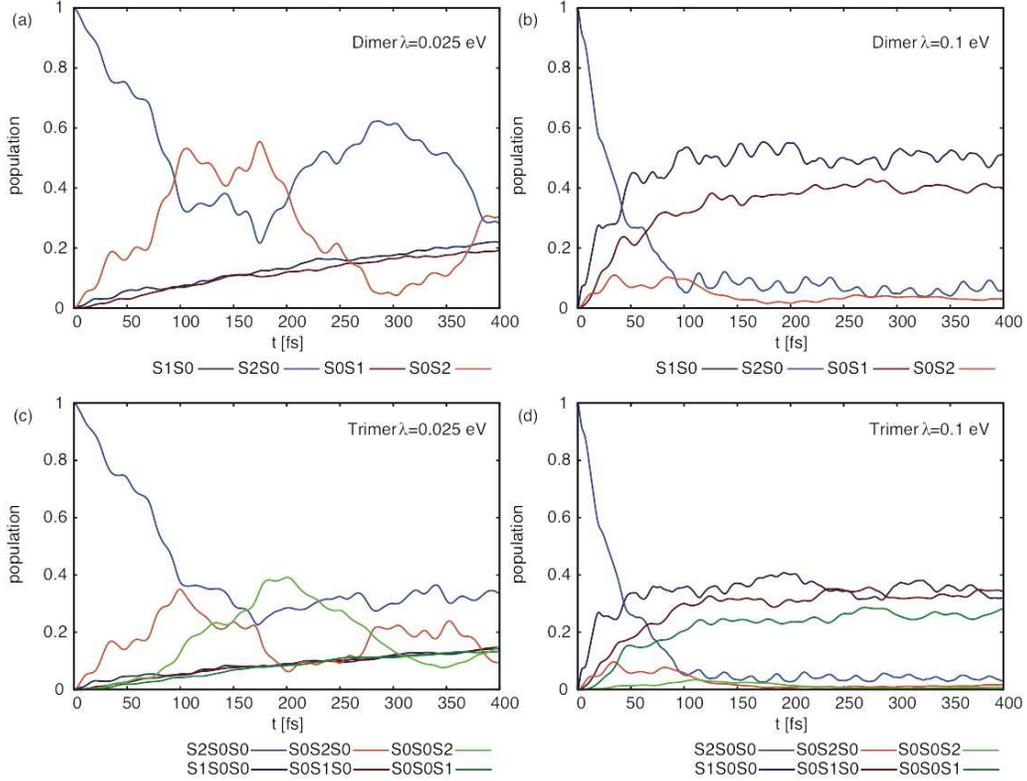}
\caption{(color online) Population dynamics of local single excitation states (see key) for  dimer (a,b) and trimer (c,d) and for two different values of the LVC constant $\lambda$. The initial excitation is to the local S$_2$ state of the first monomer ($m=1$).
}
\label{fig:pope}     
\end{center}
\end{figure}
In the following we will investigate the interplay between transfer and local deactivation in more detail. To this end we have chosen an initial condition where the dipole operator for the \Stwo tran\-si\-tions acts on monomer $m=1$ only. \fref{fig:pope} shows the subsequent population dynamics for two different values of the LVC strength. Focussing on the dimer (panel (a)) one notices that for small $\lambda$ exciton transfer between the two S$_2$ excitation state dominates, while the S$_1$ state becomes populated only slowly.  Upon increasing the LVC strength the local S$_2$ state is rapidly deactivated and the transfer proceeds in the S$_1$ manifold. This behavior carries over to the trimer in \fref{fig:pope}c,d, where the delayed population of states corresponding to excitations along the aggregate is even more visible.

The most surprising result is perhaps the behavior of the S$_1$ state populations in the weak coupling case. Their
population rise resembles that of the cases shown in \fref{fig:popdyn}c and e. This implies that the S$_1$ band is
gradually filled by the rapid energy transfer in the S$_2$ band without showing any signature of S$_1$ exciton transfer
along the aggregate  by it own.


\section{Summary} 
\label {sec:concl}
Combining models from Frenkel exciton transfer and linear vibronic coupling theory the interplay of local deactivation
processes and intermolecular energy transfer was investigated for a monomer as well as a dimer and trimer PBI system in
dependence on the  LVC strength. Although there is no obvious influence of LVC strength on low-resolution linear
absorption spectra various features appeared in high-resolution spectra as well as in population and transfer dynamics,
which can be explained by changes in the PES landscape due to the different couplings. Most notable here are  the two
opposite effects of the Coulomb coupling for a given LVC strength. First, the CIs are energetically shifted what may
favor resonance with the optically excited wave packet. Second, in the Coulomb coupled case the wave packet does not
leave the Franck-Condon region in direction of the CI, but only after backscattering from repulsive parts of the PES. The
accompanying wave packet spreading diminishes the efficiency for non-adiabatic transitions at least in the weak LVC
case.  {As net effect internal conversion becomes faster when going from the monomer to the dimer, but slows down for the
trimer. It is interesting to note that a speed-up by a factors of about ten upon dimerization has been observed in
Ref. \citenum{Kullmann:2012tf} although for a different system.}


\begin{acknowledgement}
The authors thank the Deutsche Forschungsgemeinschaft (DFG) for financial support through the Sfb 652.
\end{acknowledgement}

%
%
%
\providecommand*\mcitethebibliography{\thebibliography}
\csname @ifundefined\endcsname{endmcitethebibliography}
  {\let\endmcitethebibliography\endthebibliography}{}

\end{document}